\documentclass[aip,cha,amsmath,amssymb,preprint,author-year,author-numerical]{revtex4-1}
\usepackage{epsfig}
\usepackage{color}
\usepackage{graphicx}
\usepackage{dcolumn}
\usepackage{bm}
\input epsf

\graphicspath{{figures/}}

\begin{document}


\title{Coexisting chaotic and multi-periodic dynamics in a model of cardiac alternans}

\author{Per Sebastian Skardal}
\email{skardals@gmail.com}
\affiliation{Departament d'Enginyeria Inform\`{a}tica i Matem\`{a}tiques, Universitat Rovira i Virgili, 43007 Tarragona, Spain}

\author{Juan G. Restrepo}
\email{juanga@colorado.edu}
\affiliation{Department of Applied Mathematics, University of Colorado, Boulder, CO 80309 USA}
\date{\today}

\begin{abstract}
The spatiotemporal dynamics of cardiac tissue is an active area of research for biologists, physicists, and mathematicians. Of particular interest is the study of period-doubling bifurcations and chaos due to their link with cardiac arrhythmogenesis. In this paper we study the spatiotemporal dynamics of a recently developed model for calcium-driven alternans in a one dimensional cable of tissue. In particular, we observe in the cable coexistence of regions with chaotic and multi-periodic dynamics over wide ranges of parameters. We study these  dynamics  using global and local Lyapunov exponents and spatial trajectory correlations. Interestingly, near nodes -- or phase reversals -- low-periodic dynamics prevail, while away from the nodes the dynamics tend to be higher-periodic and eventually chaotic. Finally, we show that similar coexisting multi-periodic and chaotic dynamics can also be observed in a detailed ionic model.
\end{abstract}


\maketitle

\begin{quotation}
Sudden cardiac arrest causes over 300,000 deaths in the United States each year. This represents roughly half of all heart disease related deaths, making it the number one cause of natural death~\cite{Karma2013AR,Weiss2006CircRes}. Ventricular fibrillation, which is characterized by chaotic dynamics in heart tissue, is almost always fatal. Both experimental~\cite{Ritzenberg1984Nature,Adam1984JCE,Smith1988Circ} and theoretical~\cite{Karma1993PRL,Karma1994Chaos} studies have linked ventricular fibrillation with spatially discordant alternans, an out-of-phase period-doubling response of heart tissue to pathological conditions such as rapid pacing~\cite{Karma2007PhysicsToday,Pastore1999Circ,Watanbe2001JCE,Hayashi2007BiophysJ,Ziv2009JPhys}. In this paper we investigate numerically a continuum coupled map model for calcium-driven alternans in a one-dimensional cable derived previously by the authors~\cite{Skardal2012PRL,Skardal2014PRE} and find coexisting regions of multi-periodic and chaotic dynamics. This system provides an example of nonlocally coupled identical dynamical systems that self-organize in regions with distinct dynamical behaviors, systems which have recently attracted much attention \cite{Panaggio}.
\end{quotation}

\section{Introduction}\label{sec1}

Ventricular fibrillation (VF)--a cardiac arrhythmia that is almost always fatal--is characterized by complex spatiotemporal dynamics that are thought to be chaotic~\cite{Karma2013AR,Weiss2006CircRes}. In particular, a phenomenon known as cardiac alternans, which is characterized by beat-to-beat alternations (i.e., period 2 dynamics) in both electrical and chemical behavior~\cite{Karma2007PhysicsToday}, has been linked to VF and the onset of chaotic behavior such as spiral wave re-entry by experimental~\cite{Ritzenberg1984Nature,Adam1984JCE,Smith1988Circ} and theoretical~\cite{Karma1993PRL,Karma1994Chaos} studies. The link between alternans and VF has been strengthened by the observation that alternans can form discordantly~\cite{Pastore1999Circ,Watanbe2001JCE,Hayashi2007BiophysJ,Ziv2009JPhys}, meaning that different regions of tissue alternate out-of-phase. Spatially discordant alternans are particularly dangerous due to their tendency to promote conduction block of activity near the nodal lines that separate out-of-phase regions~\cite{Weiss2006CircRes}.

In isolation, the onset of alternans in a single cardiac cell corresponds to a period-doubling bifurcation in the beat-to-beat voltage and calcium dynamics, usually measured by the action potential duration (APD) and peak calcium concentration (Ca)~\cite{Karma2007PhysicsToday}. Importantly, this period-doubling bifurcation can be driven by an instability in either the voltage or calcium dynamics~\cite{Chudin1999BiophysJ,Shiferaw2003BiophysJ,Pruvot2004CircRes,Bien2006BiophysJ,Restrepo2008BiophysJ}. When realized in a cable or patch of tissue, the spatiotemporal dynamics of calcium-driven alternans differ qualitatively from those of voltage-driven alternans~\cite{Zhao2008PRE,Sato2006CircRes}. In particular, when alternans are calcium-driven, the length scale of phase reversals between discordant regions potentially becomes as small as the length scale of a single cell~\cite{Sato2007BiophysJ}. Recently, we developed a reduced model for the spatiotemporal dynamics of calcium-driven alternans~\cite{Skardal2012PRL,Skardal2014PRE} (henceforth referred to as the {\it SCA model} for spatiotemporal calcium alternans model), which reproduces these findings and shows that, for sufficiently large degrees of instability, calcium-driven alternans admit spatially discontinuous solutions--a class of solutions that is non--physical when alternans are voltage-driven. 

In this paper we show that for a wide range of parameters, the SCA model admits even more complex solutions. In particular, solutions can display multiple periodicities of different order and chaos--often simultaneously. For several decades, complex periodic and chaotic dynamics in cardiac tissue has been an important area of research due to its link with VF. Such dynamics have been observed in both experimental~\cite{Chialvo1990Nature,Garfinkel1997JCI} and numerical~\cite{Shiferaw2003BiophysJ} studies. Chaos has been observed in the Echebarria--Karma model~\cite{Echebarria2002PRL,Echebarria2007PRE} (an analogous reduced model for voltage-driven alternans) arising from the modulation of traveling wave patterns~\cite{Dai2010Chaos}. In contrast, we observe in the SCA model qualitatively different dynamics that are spatially localized. Such solutions can be observed in detailed ionic models such as the Shiferaw-Fox model~\cite{Shiferaw2003BiophysJ,Fox2002AJPHCP} that was used in Ref.~\cite{Skardal2014PRE} as we demonstrate below. Similar patterns consisting of bands of chaotic and regular dynamics, known as the {\it frozen random pattern}\cite{Willeboordse}, were observed originally by Kaneko\cite{Kaneko} in lattices of coupled chaotic maps. Recently there has been much renewed interest in similar types of dynamics known as {\it chimera states} in the context of coupled phase oscillators\cite{Panaggio} and more general dynamical systems \cite{Omelchenko, Schoell}. 
In this paper we show how such patterns appear in a model of spatiotemporal alternans dynamics, an example of a continuum coupled map~\cite{Ott2001PRE}. 

The remainder of this paper is organized as follows. In Sec.~\ref{sec2} we summarize the model and the bifurcations studied previously in Refs.\cite{Skardal2012PRL,Skardal2014PRE}. In Sec.~\ref{sec3} we illustrate the coexistence of chaotic and multi-periodic dynamics in the model. We support our findings by computing both global and local Lyapunov exponents and investigate the correlations of trajectories along the cable. In Sec.~\ref{sec4} we present simulations from a detailed ionic model where chaotic and multi-periodic dynamics can be easily observed. In Sec.~\ref{sec5} we conclude with a discussion of our results.

\section{Model summary}\label{sec2}

The SCA model~\cite{Skardal2012PRL,Skardal2014PRE}, which is based on the pioneering restitution-based approach of Refs.~\cite{Nolasco1968JAP,Guevara1984IEEE} and extends the amplitude equation of Refs.~\cite{Echebarria2002PRL,Echebarria2007PRE}, consists of a system of two integro-difference equations that model the beat-to-beat evolution of the non-dimensional amplitudes of calcium and voltage alternans along a one-dimensional cable\cite{foot}, assuming a calcium-mediated instability. By convention we assume that the cable has length $L$ with spatial coordinate $x\in[0,L]$ denoting the position along the cable, and that the cable is paced at the $x=0$ end with period $\tau_{BCL}$. The non-dimensional amplitude of calcium and voltage alternans at beat $n$ and location $x$ along the cable are denoted $c_n(x)$ and $a_n(x)$, respectively. Healthy period-one dynamics corresponds to $a_n(x) \equiv 0$, $c_n(x) \equiv 0$, while $a_n(x) \neq 0$ or  $c_n(x) \neq 0$ indicate alternans. In the SCA model, the beat-to-beat dynamics of $a_n(x)$ and $c_n(x)$ are governed by
\begin{widetext}
\begin{align}
c_{n+1}(x)&=-rc_n(x)+c_n^3(x)-\alpha a_n(x)+\frac{\alpha}{\Lambda}\int_0^xe^{(x'-x)/\Lambda}a_n(x')dx',\label{eq:MapC}\\
a_{n+1}(x)&=\int_0^LG(x,x')\left[-\beta a_n(x')+\frac{\beta}{\Lambda}\int_0^{x'}e^{(y-x')/\Lambda}a_n(y)dy+\gamma c_{n+1}(x')\right]dx'.\label{eq:MapA}
\end{align}
\end{widetext}
where the parameters $r$ and $\beta$ are related to the single-cell, uncoupled calcium and voltage dynamics, $\alpha$ and $\gamma$ represent the strength of voltage-to-calcium and calcium-to-voltage coupling, respectively, and $\Lambda$ is a parameter related to the restitution of conduction velocity. These parameters are summarized in Table~\ref{table:Parameters} and discussed in more detail in Appendix~\ref{appA}. The dynamics of voltage alternans [Eq.~(\ref{eq:MapA})] are spatially coupled by the Green's function $G(x,x')=G(x'-x)+G(x'+x)+G(2L-x'-x)$, where
\begin{align}
G(x) = \frac{1}{\sqrt{2\pi\xi^2}}e^{-x^2/2\xi^2}\left[1+\frac{wx}{2\xi^2}\left(1-\frac{x^2}{\xi^2}\right)\right].\label{eq:Kernel}
\end{align}
For a full derivation of the SCA model, see Refs.~\cite{Skardal2012PRL,Skardal2014PRE}. We note that the relatively simple system (\ref{eq:MapC})--(\ref{eq:Kernel}), in which each cell is described by only  two variables, reproduces nontrivial effects observed also in much more complex ionic model simulations, in which each cell is described by dozens of variables~\cite{Skardal2012PRL,Skardal2014PRE}.

 \begin{table}[t]
 \caption{\label{table:Parameters} Description of SCA model parameters given by Eqs.~(\ref{eq:MapC})--(\ref{eq:Kernel}) and the values used in this paper.}
 \begin{tabular}{ c | c | c }
 \hline
 \hline
 Parameter & Description & Value\\
 \hline
 $r$ & degree of calcium instability & varied \\
 $\Lambda$ & slope of conduction velocity restitution & $30$ \\
 $\beta$ & slope of APD restitution & $0$ \\
 $\alpha$ & voltage $\to$ calcium coupling & $\sqrt{0.3}$ \\
 $\gamma$ & calcium $\to$ voltage coupling & $\sqrt{0.3}$ \\
 $\xi$ & length scale of electronic coupling & $1$ \\
 $w$ & asymmetry of electronic coupling & $0$ \\
 \hline
 \hline
 \end{tabular}
 \end{table}

For simplicity, in this paper we focus on the effects of changing the main dynamical parameter $r$. To understand the role of this parameter, note that in the absence of voltage alternans, $a_n = 0$ (such as when using a voltage clamp), the single-cell dynamics of calcium alternans are modeled by
\begin{align}
c_{n+1}=-rc_n+c_n^3,\label{eq:LocalC}
\end{align}
for which the no alternans solution ($c_n=0$) is stable for $0\le r<1$. At $r=1$ this solution loses stability and gives rise to stable non-zero solutions for $r>1$. The parameter $r$ thus can be interpreted as controlling the degree of instability in the calcium cycling machinery of the cell. From now on, we will study the effects of increasing $r$ while keeping the other parameters constant.
 
The dynamics of the SCA model for relatively low values of $r$ was studied in Refs.~\cite{Skardal2012PRL,Skardal2014PRE}. As $r$ is increased from zero three types of dynamics are observed: no alternans, smooth wave patterns, and discontinuous patterns. The no alternans solution, stable for sufficiently small $r$, is given by $c(x),a(x)\equiv0$. The first bifurcation, corresponding to the onset of alternans, separates the no alternans solutions from the smooth wave pattern solutions. Smooth wave pattern solutions can be either stationary or have a finite velocity, depending on whether the asymmetry of the Green's function [controlled by the parameter $w$ in (\ref{eq:Kernel})] is large or small, respectively. In the case of small or no asymmetry, as we consider here, smooth wave patterns have a finite velocity with which they move towards the pacing site. Finally, at a second bifurcation the smooth wave patterns give way to solutions where the calcium profiles form discontinuous jumps at each phase reversal while the voltage profiles remain smooth. Furthermore, these solutions are always stationary. In the remainder of this paper we will see that the SCA model admits even more complex dynamics.

\section{Coexisting chaotic and multiperiodic dynamics}\label{sec3}

We will now study the dynamics of the SCA model for even larger values of the calcium-instability parameter $r$. In particular, we find a wide parameter range that admits simultaneous multi-periodic and chaotic behavior that is spatially localized. We will study these dynamics over a large range of $r$ values while keeping all other parameters fixed. In particular, throughout this paper we use $\Lambda=30$, $\alpha,\gamma=\sqrt{0.3}$, $\beta=0$, $\xi=1$, and $w=0$ and consider a cable of length $L=20$. Furthermore, we will consider initial conditions defined randomly, drawing each point $c_0(x)$ uniformly from $[0.68,1.08]$ if $6\le x<16$, and otherwise draw $c_0(x)$ from $[-1.08,-0.68]$. This choice is made to ensure the presence of two nodes along the cable, one at $x=6$ and the other at $x=16$. We emphasize here that dynamics similar to those we will present can be observed for other choices of parameters and initial conditions. Finally, to update Eqs.~(\ref{eq:MapC}) and (\ref{eq:MapA}), we evaluate each integral by discretizing the interval $[0,20]$ using $\Delta x = 0.004$ and using the trapezoidal rule.

\subsection{Chaos and multiple periodicities}

\begin{figure*}[t]
\centering
\epsfig{file =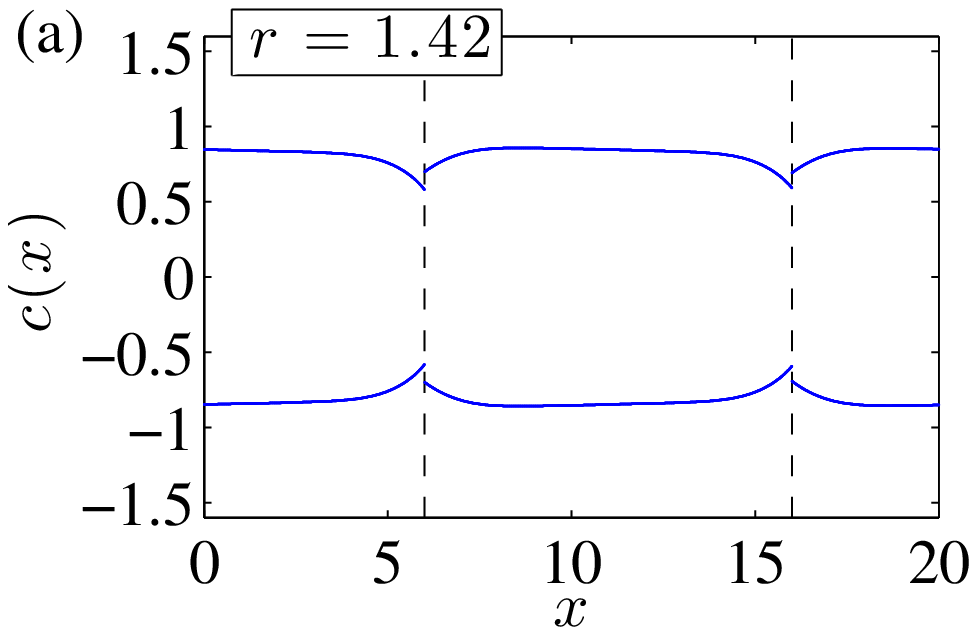, clip =,width=0.31\linewidth }
\epsfig{file =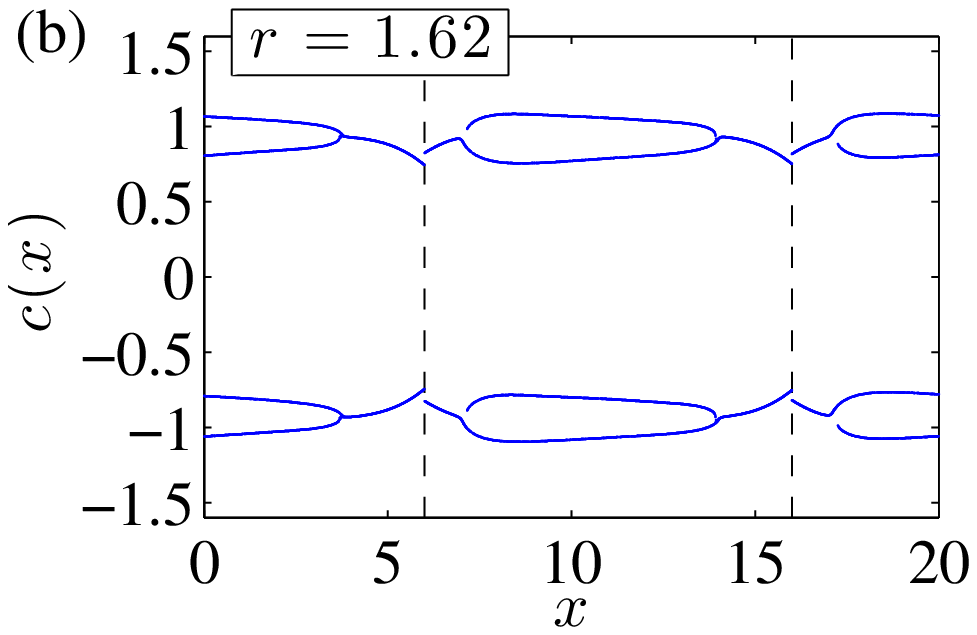, clip =,width=0.31\linewidth }
\epsfig{file =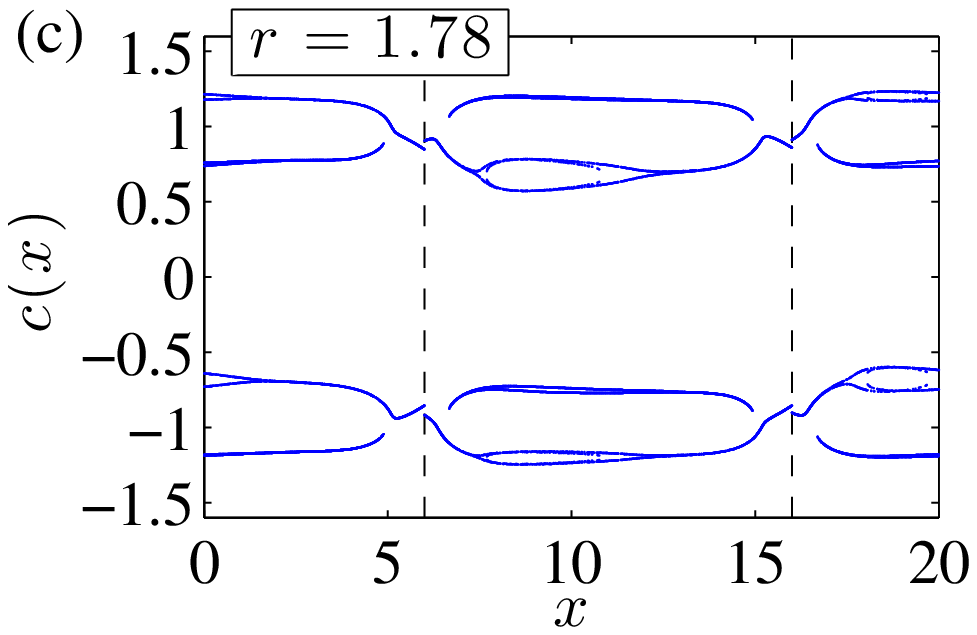, clip =,width=0.31\linewidth }\\
\epsfig{file =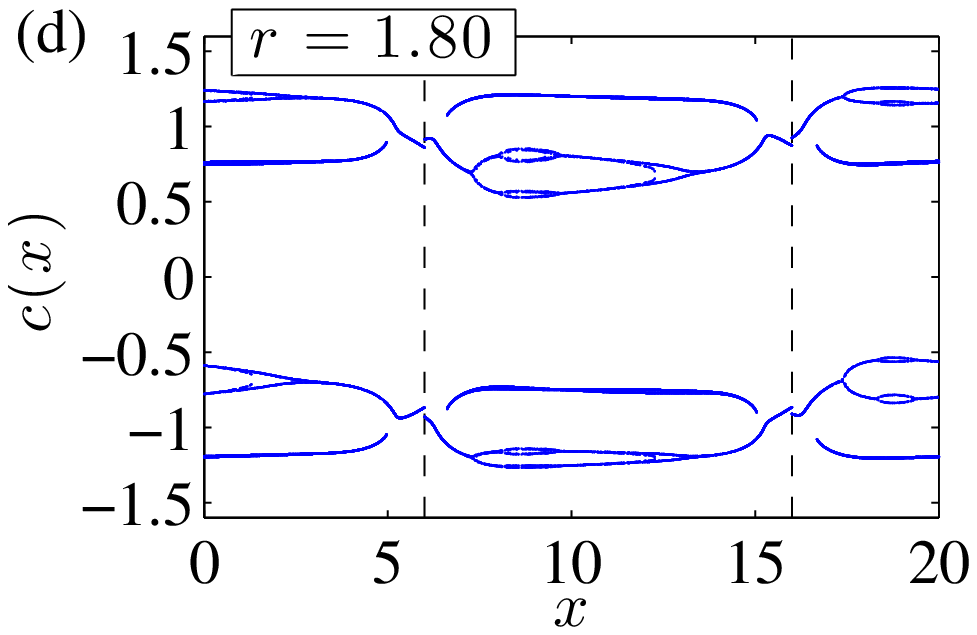, clip =,width=0.31\linewidth }
\epsfig{file =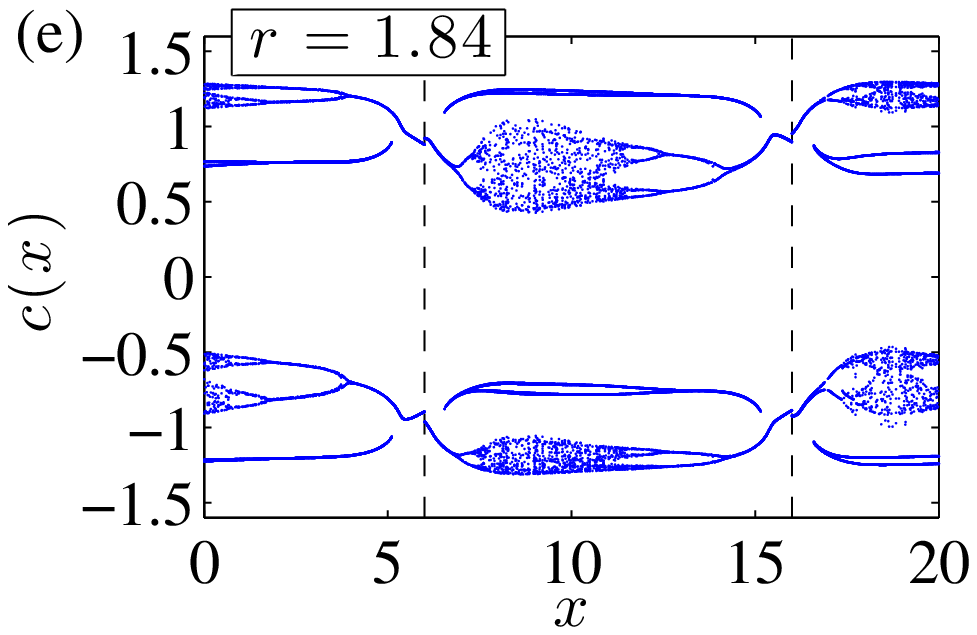, clip =,width=0.31\linewidth }
\epsfig{file =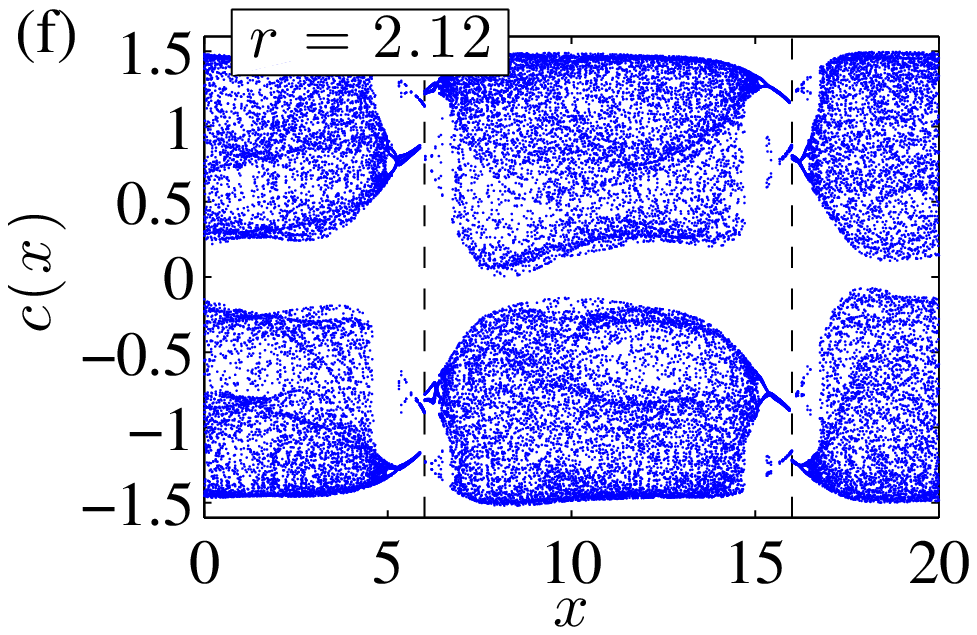, clip =,width=0.31\linewidth }
\caption{(Color online) Coexisting chaotic and multi-periodic dynamics: Example steady-state calcium profiles $c(x)$ obtained from direct simulation of Eqs.~(\ref{eq:MapC}) and (\ref{eq:MapA}) for (a) $r=1.42$, (b) $r=1.62$, (c) $1.78$, (d) $1.80$, (e) $1.84$, and (f) $2.12$. Steady-state is reached after a long transient of $2000$ beats. Phase reversals at $x=6$ and $x=16$ are marked by vertical dashed lines.}\label{fig:Ex}
\end{figure*}

We begin by presenting in Fig.~\ref{fig:Ex} evidence from direct numerical simulations of Eqs.~(\ref{eq:MapC}) and (\ref{eq:MapA}) of coexisting chaotic and multi-periodic dynamics. In panels (a)--(f) we plot  $c_n(x)$ for $n = 2001$ to $n = 2016$ after discarding the initial $2000$ beats for several values of $r$: $r=1.42$, $1.62$, $1.78$, $1.80$, $1.84$, and $2.12$. Thus, the plots show an approximation to an attractor for each value of $x$. For purposes of visualization we mark the locations of the phase reversals at $x=6$ and $x=16$ with dashed vertical lines. These profiles serve as good examples for the increasingly complex behavior we observe for larger $r$ values.

For $r=1.42$ the steady-state dynamics are period-two and fall into the category of solutions studied in Refs.~\cite{Skardal2012PRL,Skardal2014PRE}. Next, at $r=1.62$ we see that a section of $c_n(x)$ away from the phase reversals has undergone a bifurcation. We note, however, that the dynamics in these regions are not period-four, but represent two separate branches of period-two dynamics that are each realized by roughly half the points along the cable due to the random initial conditions. This interesting effect can be explained using the local calcium map in Eq.~(\ref{eq:LocalC}) and will be discussed below. We note that other branchings that we observe are in fact the result of period-doubling bifurcations. In particular, for $r=1.78$ and $r=1.80$ we observe solutions with high-order periodicities away from the phase reversals but which remain period-two near the phase reversals. Finally, for $r=1.84$ and $r=2.12$ we observe chaotic behavior. At $r=1.84$ the chaos is localized to relatively small parts of the cable away from the phase reversals, with multi-periodic behavior elsewhere. Finally, chaos dominates for $r=2.12$, with only small areas of periodic behavior present near the phase reversals.

In addition to different types of dynamics coexisting in different regions, different attractors can coexist in the same region. More specifically, we find that in some regions of the cable there can be multiple attractors which $c_n(x)$ can approach as $n\to \infty$ depending on the initial conditions. One example is in Fig.~\ref{fig:Ex} (b) in the region $7 \lesssim x \lesssim 13$. In this region there are two period-two orbits, so that, for a fixed value of $x$, $c_n(x)$ alternates only between two values. However, which periodic orbit $c_n(x)$ approaches depends sensitively on the initial conditions, and in our case each of the two periodic orbits is approached at roughly half the values of $x$. Therefore the plot seems to show four curves even though each point alternates at most between two values. Another example occurs in Fig.~\ref{fig:Ex} (e) for $8 \lesssim x \lesssim 10$ where a chaotic attractor coexists with another which, although apparently periodic, is upon close inspection (not shown) also chaotic.

\subsection{Local map dynamics}

\begin{figure}[t]
\centering
\epsfig{file =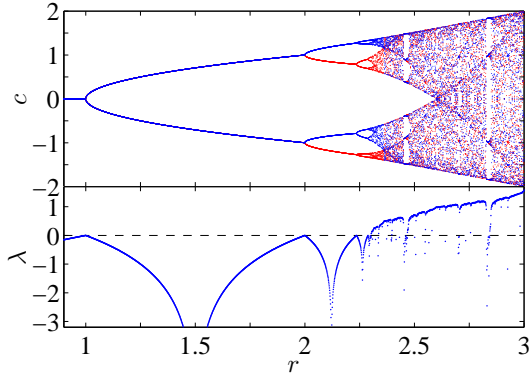, clip =,width=0.48\linewidth }
\caption{(Color online) Local map dynamics: Bifurcation diagram of steady-state solutions and Lyapunov exponent for the one-dimensional local map given by Eq.~(\ref{eq:LocalC}).}\label{fig:Map}
\end{figure}

To gain some insight into the dynamics we have observed, we examine in more detail the dynamics described by the one-dimensional map in Eq.~(\ref{eq:LocalC}) as a function of the parameter $r$. For $r\in[0,3]$, Eq.~(\ref{eq:LocalC}) maps the interval $[-2,2]$ into itself. A straightforward analysis shows that for $r\in[0,1)$ the fixed-point solution $c=0$ is stable and loses stability at $r_1=1$. Immediately above $r_1$, the period-two solution $c=\pm\sqrt{r-1}$ is stable, but then loses stability at $r_2=2$. Interestingly, instead of a single period-four solution, two stable period-two solutions are born whose basins of stability partition the interval $[-2,2]$. This bistability is due to the presence of two extremal points in the cubic map and is similar to the dynamics found in other one-dimensional maps~\cite{Glass1982PRL,Perez1982PLA}. While bistability between periodic and chaotic dynamics is observed in these other maps, the symmetry in Eq.~(\ref{eq:LocalC}) implies that both stable solutions are either both periodic or both chaotic. These solutions, which are given by $c=\pm\sqrt{r\pm\sqrt{r^2-4}}/\sqrt{2}$, are themselves stable until $r_3=\sqrt{5}$, when two period-four solutions are born. As with many other one-dimensional maps, this process of period-doubling branchings continue as a cascade until the onset of chaotic behavior~\cite{Strogatz2001}.

In Fig.~\ref{fig:Map} we illustrate the dynamics of the local map with its bifurcation diagram. We highlight the splitting of stable periodic solutions at $r_2=2$ by plotting one family in blue and the other in red. Note, however, that solutions are symmetric about $c=0$. We also compute for each $r$ value the Lyapunov exponent defined as
\begin{align}
\lambda = \lim_{n\to\infty}\frac{1}{n}\log_2\frac{|\delta c_n|}{|\delta c_0|},\label{eq:Lyapunov1}
\end{align}
where $\delta c_0$ is formally an infinitesimal perturbation to a solution in the attractor and $\delta c_n$ is the evolved perturbation after $n$ iterations, such that $\lambda$ describes the rate of divergence (or convergence) of two nearby trajectories. In Fig.~\ref{fig:Map} we plot a numerical approximation to $\lambda$ (bottom panel) below $c$ (top panel) as a function of $r$. From both the bifurcation diagram and the Lyapunov exponent we find that the onset of chaotic behavior occurs at $r_c\approx2.303$.

\subsection{Local and global Lyapunov exponents}

We now return to the full system given by Eqs.~(\ref{eq:MapC}) and (\ref{eq:MapA}). Our objective in this Section is to quantify the separation of the cable into distinct chaotic and periodic regions. In order to do this, we define a {\it local} Lyapunov exponent $\lambda_{\text{local}}(x)$ for every point $x$ in the cable. For the purposes of this paper, we will define the dynamics at a point to be chaotic at a point $x$ if $\lambda_{\text{local}}(x) > 0$. In addition to the local Lyapunov exponent, we will also consider a global Lyapunov exponent. To define these exponents, we consider an infinitesimal perturbation $\delta c_0(x)$ to a solution $c(x)$ in the attractor. If $\delta c_n(x)$ denotes the evolution of the perturbation forward in time $n$ steps, we define the local Lyapunov exponent as
\begin{align}
\lambda_{\text{local}}(x)=\lim_{n\to\infty}\frac{1}{n}\log_2\frac{|\delta c_n(x)|}{|\delta c_0(x)|}.\label{eq:LyapunovLocal}
\end{align}
Importantly, $\lambda_{\text{local}}(x)$ depends on $x$ and therefore allows us to compare the dynamics at different points along the cable. We can calculate also the global Lyapunov exponent, denoted $\lambda_{\text{global}}$, as
\begin{align}
\lambda_{\text{global}}=\lim_{n\to\infty}\frac{1}{n}\log_{2}\frac{\|\delta c_n(x)\|}{\|\delta c_0(x)\|},\label{eq:LyapunovGlobal}
\end{align}
where $\|\cdot\|$ represents the $L^2$--norm, i.e.,
$\|c(x)\|=\sqrt{\int_0^Lc^2(x)dx}.\label{eq:L2Norm}$
Thus, $\lambda_{\text{global}}$ gives a single value that describes the aggregate dynamics of $c_n(x)$ over the whole cable. (Note that, since the choice of $\delta c_0(x)$ is arbitrary, $\lambda_{\text{global}}$ corresponds with probability one to the largest Lyapunov exponent.) 

\begin{figure*}[t]
\centering
\epsfig{file =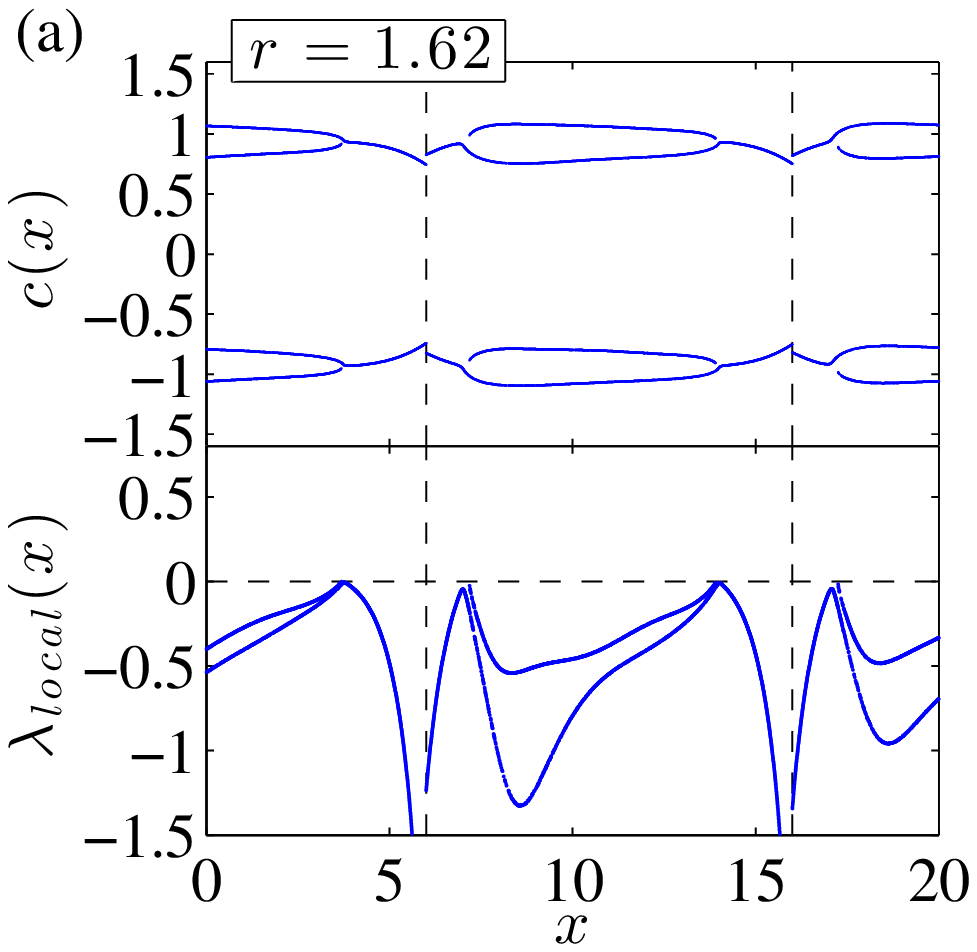, clip =,width=0.31\linewidth }
\epsfig{file =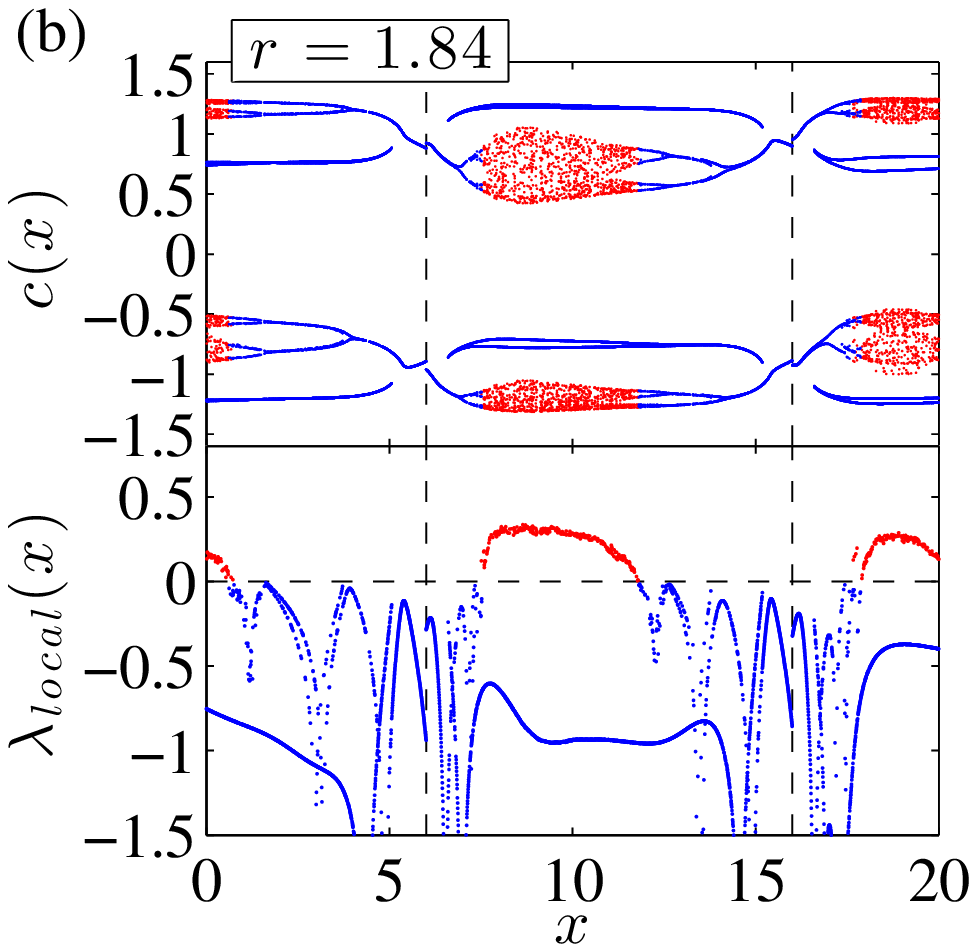, clip =,width=0.31\linewidth } 
\epsfig{file =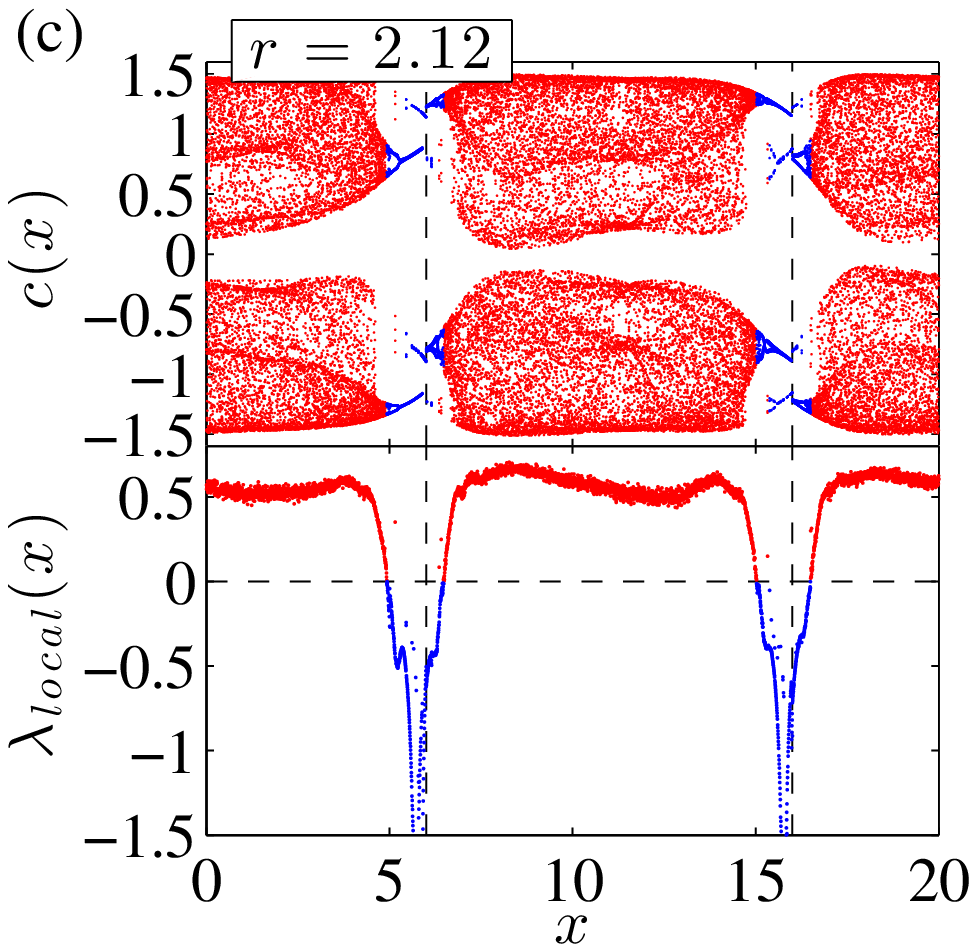, clip =,width=0.31\linewidth } 
\caption{(Color online) Local Lyapunov exponent: Steady-state calcium profiles $c(x)$ with corresponding spatial Lyapunov exponents $\lambda_{\text{local}}(x)$ for (a) $r=1.62$, (b) $r=1.84$, and (c) $r=2.12$. Points displaying chaotic behavior, characterized by $\lambda_{\text{local}}(x)>0$, are colored red.}\label{fig:LyapLocal}
\end{figure*}

In order to quantify the coexistence of chaos and regular behavior observed in Fig.~\ref{fig:Ex}, we plot in Fig.~\ref{fig:LyapLocal} (a)--(c) the $c(x)$ attractors and the corresponding Lyapunov exponent $\lambda_{\text{local}}(x)$ for $r=1.62$, $r=1.84$, and $r=2.12$. Using $\lambda_{\text{local}}(x)$, we identify which points along the cable display chaotic dynamics by checking if $\lambda_{\text{local}}(x)>0$ and indicate these areas of the cable by coloring both $c(x)$ and $\lambda_{\text{local}}(x)$ red. Otherwise, we color both $c(x)$ and $\lambda_{\text{local}}(x)$ blue. We remark that, as discussed before, for some values of $x$ there are multiple coexisting attractors, and correspondingly there are some regions that have multiple curves [e.g., around $x=10$ in panels (a) and (b)]. 

For $r=1.62$ no chaotic behavior is exhibited anywhere in the cable. However, we observe that $\lambda_{\text{local}}(x)$ tends to zero at the points corresponding to the branchings, analogous to branchings of typical bifurcation diagrams (e.g., Fig.~\ref{fig:Map}), and $\lambda_{\text{local}}(x)$ is most negative close to the nodes at $x = 6$ and $x = 16$. For $r=1.84$ and $r=2.12$ the chaotic regions as indicated by $\lambda_{\text{local}}(x)>0$ agree with what one would expect from observing the upper panels in Fig.~\ref{fig:LyapLocal}. For $r=1.84$ chaotic behavior is limited to a relatively small fraction of the cable, while for $r=2.12$ it covers most of the cable.

\begin{figure}[t]
\centering
\epsfig{file =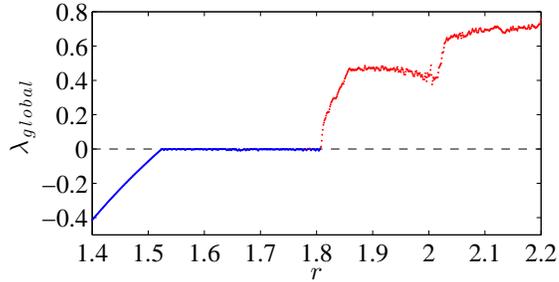, clip =,width=0.48\linewidth }
\caption{(Color online) Global Lyapunov exponent: Global Lyapunov exponent $\lambda_{\text{global}}$ computed over a range of $r$ values.}\label{fig:LyapGlob}
\end{figure}

To check that the full spatiotemporal system is also chaotic, we calculate the global Lyapunov exponent $\lambda_{\text{global}}$ as function of $r$, and plot it in Fig.~\ref{fig:LyapGlob}. Here we see that, on aggregate, the dynamics transition from the non-chaotic regime ($\lambda_{\text{global}}<0$) at smaller $r$ values, to the chaotic regime ($\lambda_{\text{global}}>0$) at larger $r$ values. We checked that the range of intermediate $r$ values (approximately between $r=1.52$ and $1.81$) where $\lambda_{\text{global}}$ remains very close to zero corresponds to parameter values that yield multi-periodic behavior with one or more branchings along the cable. We also note that $\lambda_{\text{global}}$ become positive when the first local chaotic regions in the cable appear.

\subsection{Trajectory correlations}

We have found that at a given point $x$ in the cable the dynamics of $c_n(x)$ can have sensitive dependence on initial conditions, suggesting chaotic dynamics. However, given the nonlocal coupling present in Eqs. (\ref{eq:MapC})-(\ref{eq:MapA}), it is unclear whether the dynamics at different locations are correlated. To investigate this further, we introduce the marginal and joint occupation probabilities $P_x(c)$ and $P_{xy}(c,c')$. In particular, $P_x(c)dc$ gives the steady-state probability that $c_n(x)$ is between $c$ and $c+dc$, while $P_{xy}(c,c')dcdc'$ is the steady-state probability that $c_n(x)$ and $c'_n(y)$ are simultaneously between $c$ and $c+dc$ and $c'$ and $c'+dc'$, respectively. We are primarily concerned with solutions where chaos dominates the cable, so as an example we restrict our attention to the parameter value $r=2.12$ [see Fig.~\ref{fig:Ex}(f)]. 

\begin{figure}[t]
\centering
\epsfig{file =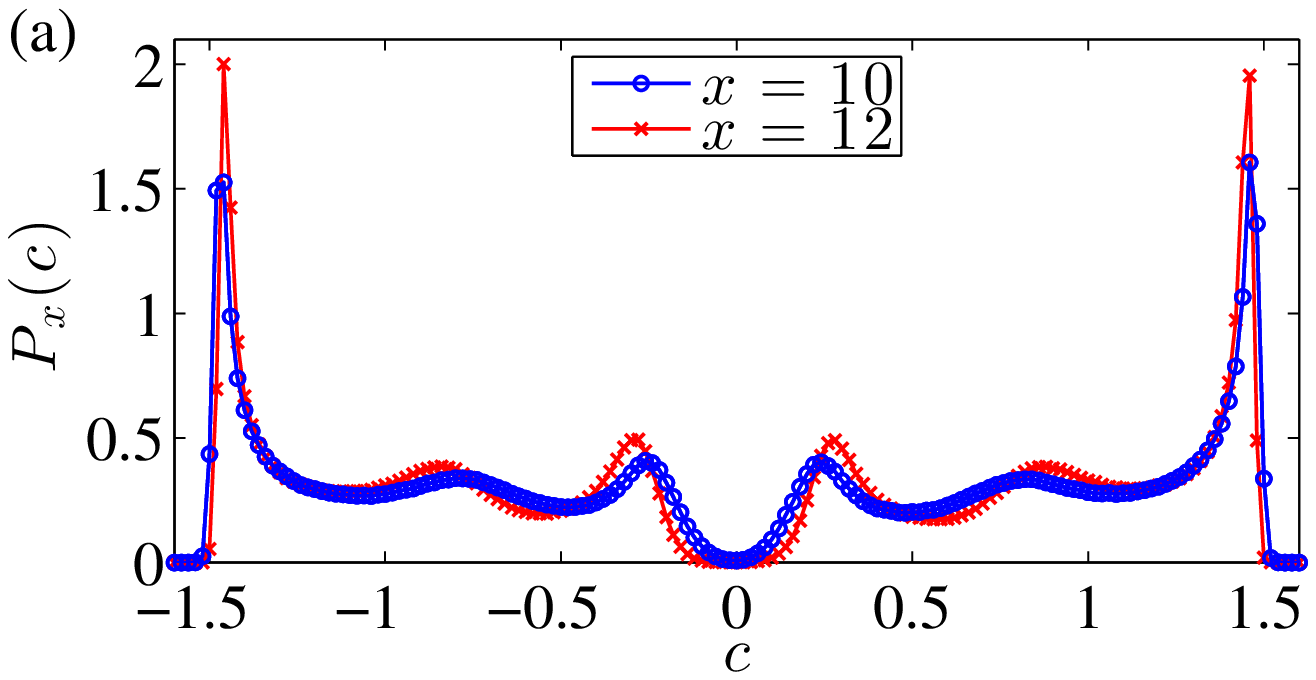, clip =,width=0.48\linewidth }
\epsfig{file =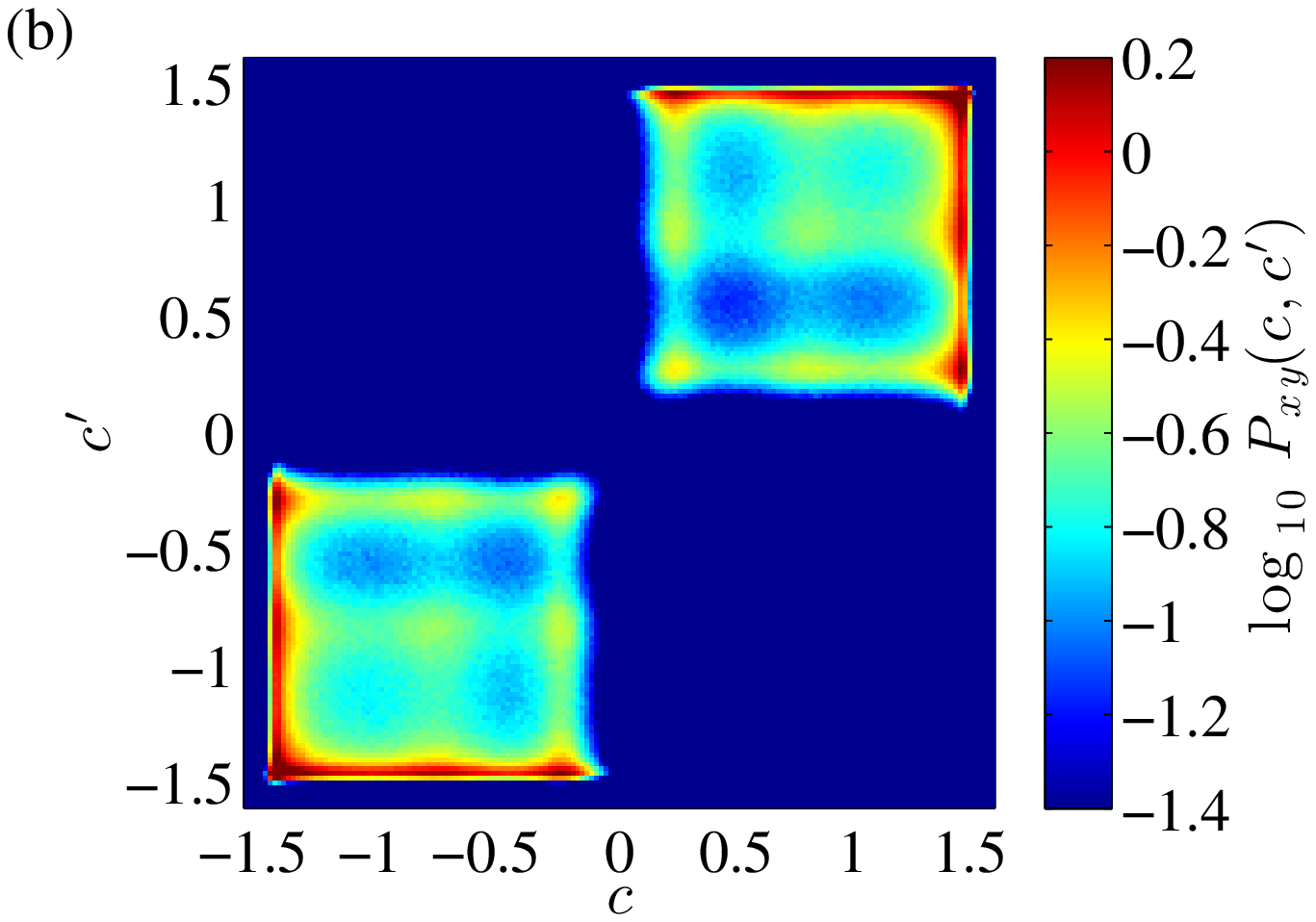, clip =,width=0.48\linewidth }
\caption{(Color online) Trajectory occupation probabilities: (a) Marginal distribution $P_x(c)$ and (b) joint distribution $P_{xy}(c,c')$ of chaotic trajectories in the set at $x=10$ and $y=12$ for $r=2.12$.}\label{fig:P}
\end{figure}

We begin by computing the marginal and joint probabilities distributions $P_x(c)$ and $P_{xy}(c,c')$ at points $x=10$ and $y = 12$ along the cable. In practice, we iterate the map (\ref{eq:MapC})-(\ref{eq:MapA}), sampling $10^7$ iterations after discarding the initial $2000$ steps, and calculating each distribution from the fraction of iterations that fall into the appropriate bins of size $dc,dc'=0.02$. In Fig.~\ref{fig:P}(a) and (b) we plot the marginal distribution $P_x(c)$ for $x=10$ (blue circles) and $12$ (red crosses) and the joint distribution $P_{xy}(c,c')$ for $x=10$ and $y=12$, respectively. Inspecting the marginal distributions $P_x(c)$ in Fig.~\ref{fig:P}(a) first, the occupation probabilities computed at $x=10$ and $12$ have similar but slightly different shape. In particular, for $x=12$ the gap about $c=0$ is wider and the peaks are larger. Next, we observe some strong structural correlations in the joint distribution $P_{xy}(c,c')$. In particular, the support of $P_{xy}(c,c')$ lies solely in the first and third quadrants, with $P_{xy}(c,c')=0$ whenever $c\cdot c'<0$. This effect comes directly from the fact that both points $x$ and $y$ were chosen from the same in-phase region along the cable. If, on the other hand, $x$ and $y$ are chosen on opposite sides of a node (e.g., $x=10$ and $y = 18$), then the joint distribution flips such that its support lies solely in the second and fourth quadrants, with $P_{xy}(c,c')=0$ whenever $c\cdot c'>0$ (not shown).

This strong structural effect present in the joint distribution implies that the occupation probabilities at different points along the cable are correlated [i.e., the joint distribution cannot be separated into the product of the two marginal distributions, $P_{xy}(c,c')\ne P_x(c)P_y(c')]$. We now investigate whether any further correlations exist past the shared/opposite sign structure that is due to the switching inherent in the system. To this end, we construct from the time series $c_n(x)$ a new time series $\tilde{c}_n(x)=|c_n(x)|$ and introduce the corresponding marginal and joint distributions $\tilde{P}_x(\tilde{c})$ and $\tilde{P}_{xy}(\tilde{c},\tilde{c}')$ analogous to those introduced above. Using the same technique, we compute the marginal and joint distributions at the same points along the cable. 
We observe (not shown) that the two distributions are very similar, with the main difference being that $\tilde{P}_{xy}(\tilde{c},\tilde{c}')$ is coarser; an effect that is simply due to sampling the joint distribution. To confirm this, we calculate the Pearson correlation coefficient of the two time series, given by $\rho=[E(\tilde c \tilde c')-E(\tilde c)E(\tilde c)]/[\sigma(\tilde c)\sigma(\tilde c')]$, where $E(\cdot)$ and $\sigma(\cdot)$ denote expected value and standard deviation, respectively. We find that the correlation coefficient is very small, $\rho=2.3\times10^{-4}$, confirming that correlations are effectively zero. We have checked that the correlation coefficient remains as small as long as $x \neq y$ and both $x$ and $y$ are both in a chaotic region. Thus, even though the dynamics has a coherent large scale structure [e.g., Fig.~\ref{fig:Ex} (e) and (f)] the chaotic behavior at two different points can be considered statistically independent. 

\section{Chaotic and Multi-periodic Dynamics in a Detailed Ionic Model}\label{sec4}

We now turn our attention to briefly study the dynamics of a detailed ionic model. Specifically, we will demonstrate that the chaotic and multi-periodic dynamics observed and studied in the reduced model above can also be observed in more complicated, biologically robust models. We consider here the Shiferaw-Fox ionic model, which combines the calcium cycling dynamics of Shiferaw et al.~\cite{Shiferaw2003BiophysJ} with the ionic current dynamics of Fox et al.~\cite{Fox2002AJPHCP}. Importantly, the coupling between detailed calcium and voltage dynamics given by the Shiferaw-Fox model allows for a robust enough model to produce calcium-driven alternans for relatively large parameter ranges.

We note that the choice of parameters and implementation we use here is the same as used in Refs.~\cite{Skardal2014PRE,Sato2007BiophysJ,KroghMadsen2007BiophysJ} except when indicated. In the calcium-cycling dynamics of the Shiferaw-Fox model, the primary mechanism for calcium ions entering the cell cytoplasm, aside from the standard L-type calcium current, is the release of stored calcium from the sarcoplasmic reticulum (SR), a network of rigid tubule-like structures that store calcium within the cell. This release occurs via a positive-feedback process in response to the activation of the L-type calcium current. In the Shiferaw-Fox model, the rate of calcium release by this mechanism is determined by a parameter $u$, where large (small) values typically correspond to more (less) instability in the calcium cycling dynamics. To promote calcium instabilities, we choose a relatively large release parameter of $u=30$ ms$^{-1}$.  In addition, we use a relatively small voltage inactivation timescale, $\tau_f=38$ ms, to ensure that voltage dynamics do not drive the instability. To ensure that calcium-to-voltage coupling is positive, we choose a relatively small calcium inactivation exponent $\gamma=0.5$~\cite{Shiferaw2003BiophysJ}. Finally, we also increase the timescale $\tau_j$ of the fast-sodium $j$-gate dynamics. This effectively increases the slope of the conduction velocity restitution curve (see Refs.~\cite{Sato2006CircRes, Skardal2014PRE,KroghMadsen2007BiophysJ} for a discussion). Here we used $\tau_j\mapsto 8\tau_j$. In summary, our parameter choices are made so that (i) alternans are calcium-driven, (ii) calcium-to-voltage coupling is positive, and (iii) the conduction velocity restitution curve is not flat.

\begin{figure}[t]
\centering
\epsfig{file =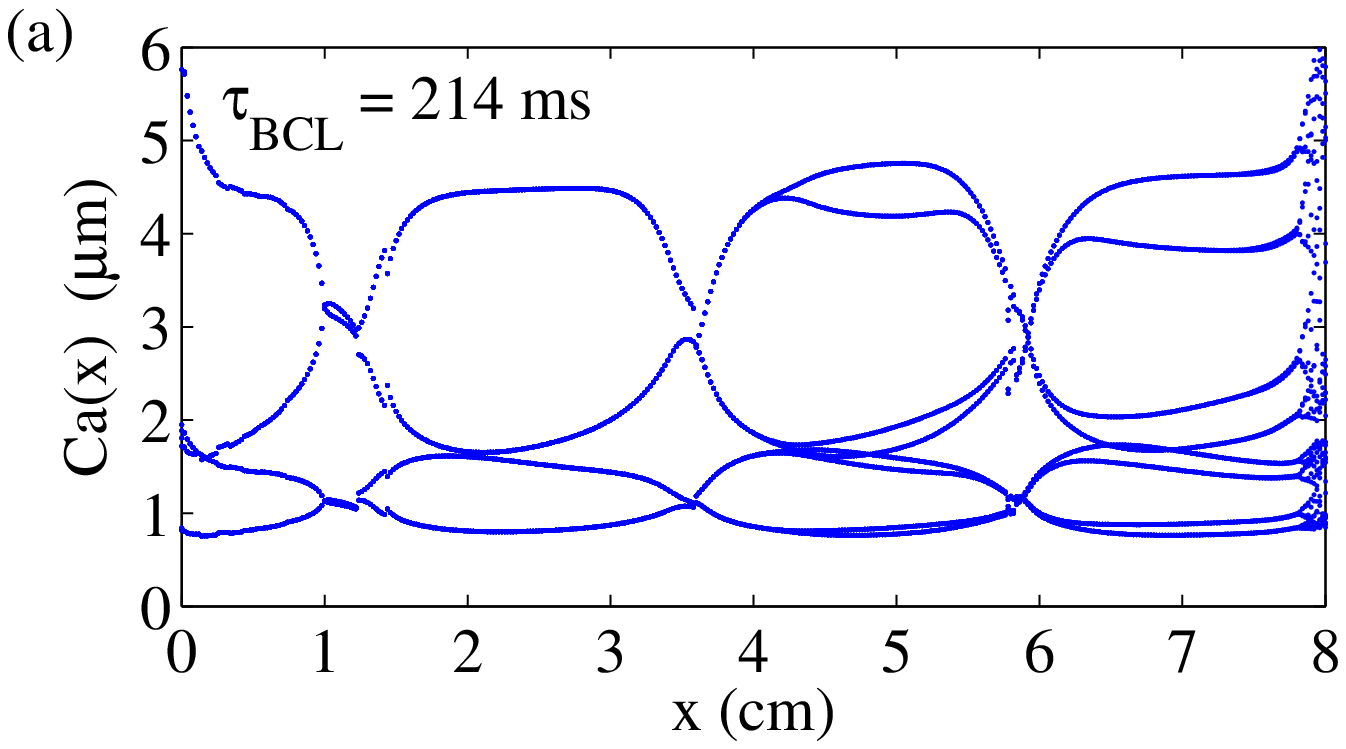, clip =,width=0.48\linewidth }
\epsfig{file =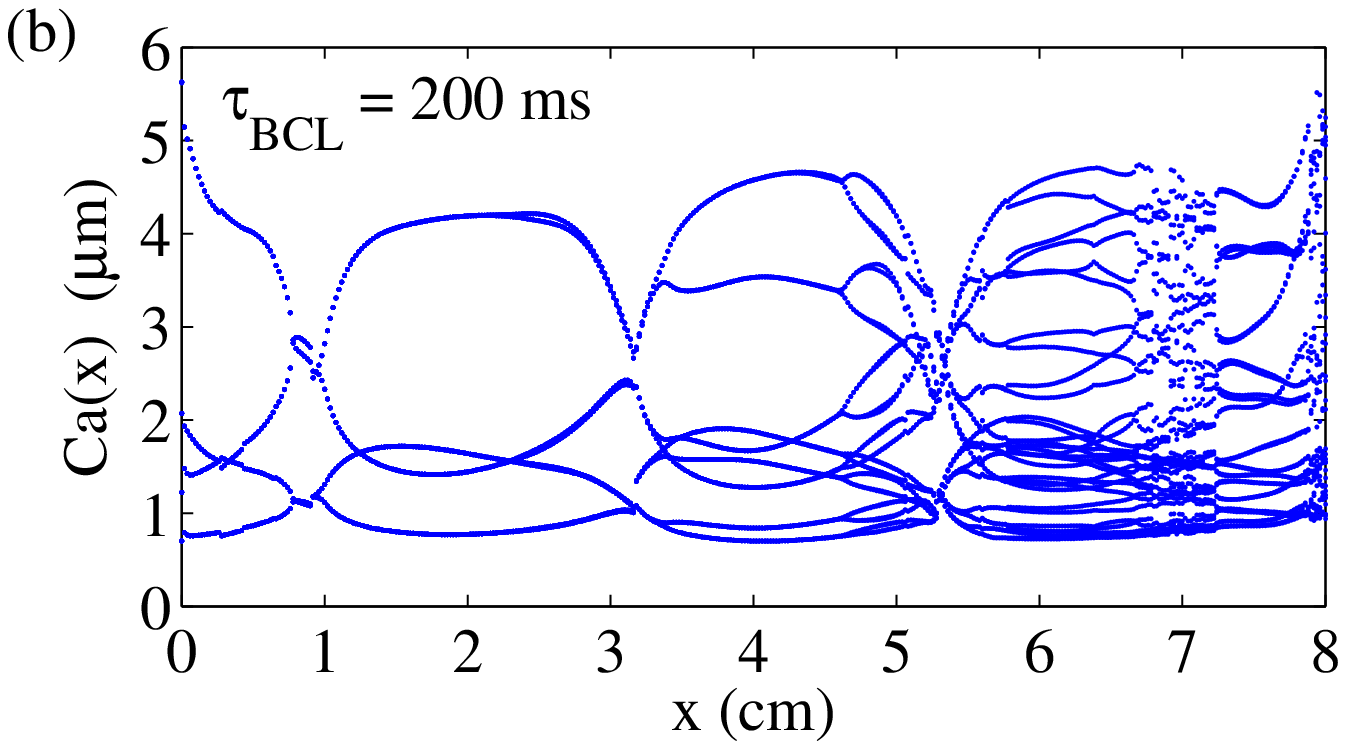, clip =,width=0.48\linewidth }
\caption{(Color online) Ionic model dynamics: Steady-state dynamics of the Shiferaw-Fox ionic model on a cable of length $8$ cm paced at $\tau_{BCL}=214$ ms (a) and $200$ ms (b). Shown are the last $32$ beats after a transient of $2000$ beats in each case.}\label{fig:IonicModel}
\end{figure}

We now present the results from simulations of the Shiferaw-Fox model. We consider here a cable of length $8$ cm using a discretization of $\Delta x=0.02$ cm paced periodically at the end $x=0$. In Fig.~\ref{fig:IonicModel} we plot the steady-state peak calcium concentration $Ca(x)$ along the cable taken from the last $32$ beats after a transient of $2000$ beats for simulations paced at $\tau_{BCL} = 214$ ms (a) and $200$ ms (b). At $\tau_{BCL}=214$ the dynamics along the cable are period-two near the nodes and period-four and period-eight away from the nodes. (We note that these are truly period-four solutions, not two different period-two solutions.) At $\tau_{BCL}=200$ ms the dynamics become even more complicated. While low periodic behavior still prevails near the nodes, we find additionally segments of the cable that have period-sixteen, period-thirty two, and chaotic dynamics. Interestingly, the more complicated dynamics, i.e., higher periodicities and chaos, tend to occur towards the back end of the cable. Also, starting at period-four, the dynamics in the top- and bottom-half branches cross one another. Nonetheless, our results confirm that the coexistence of multi-periodic and chaotic dynamics we have studied in the reduced model above [Eqs.~(\ref{eq:MapC}) and (\ref{eq:MapA})] is not just an artifact, but can be realized in a biologically robust ionic model.

\section{Discussion}\label{sec5}

In this paper we have numerically studied novel dynamics that emerge from the SCA model for calcium-driven alternans in a one dimensional cable of tissue~\cite{Skardal2012PRL,Skardal2014PRE}. In particular, we have observed spatially localized chaotic and multi-periodic behavior that often occurs simultaneously over a wide range of parameters. To study these dynamics we have used both local and global Lyapunov exponents, as well as investigated the occupation probabilities and the correlations between them for chaotic trajectories. Interestingly, as the degree of calcium instability increases, the dynamics away from phase reversals tends to be more complicated, resulting in high-order periodicity and eventually chaos, while dynamics near the node tends to be simpler, often period-two. 

We have complemented our analysis of the reduced model with numerical simulations of the Shiferaw-Fox ionic model, a biologically robust model that has a detailed description of the intracellular calcium cycling dynamics. In particular, we showed that coexisting multi-periodic and chaotic dynamics can be observed in a cable for a reasonable set of parameters.

Theoretical efforts to eliminate alternans by implementing control algorithms are generally focused on the suppression of a relatively small number of unstable modes in the weakly nonlinear regime \cite{Control}. In the strongly nonlinear regime analyzed here, the vanishing correlation length that we observed suggests that chaos is of very high dimensionality and that these methods would not be effective. However, we note that the parameters in which localized chaos are present are somewhat extreme. 

Our work shows that the SCA model is an example of a system of non-locally coupled dynamical systems that, despite being defined identically, self-organize into spatially localized regions with distinct dynamical behavior. These kinds of systems have recently attracted much attention, in particular in systems of coupled oscillators, where they have become known as chimera states~\cite{Panaggio}. Our results, following Refs.\cite{Omelchenko, Schoell}, suggest that the idea of a chimera state can be generalized to a larger class of dynamical systems that display much wider ranges of dynamical behavior than just synchronization and incoherence. Finally, we believe that the possibility of such dynamics possibly existing in physically relevant models of cardiac tissue is interesting for the cardiac dynamics community, and potentially relevant to the design of alternans control protocols.

\acknowledgments

The work of P.S.S. was supported by the James S. McDonnell Foundation.

\begin{appendix}

\section{Model parameters}\label{appA}

In this appendix section we briefly describe the parameters of the SCA model given by Eqs.~(\ref{eq:MapC})--(\ref{eq:Kernel}). For more detail on the derivation of the SCA model and the parameters, see Refs.~\cite{Skardal2012PRL,Skardal2014PRE}. In Table~\ref{table:Parameters} we summarize all model parameters.
 
Recall that the main parameter varied in this paper is $r$, which controls the single cell calcium dynamics as described in Eq.~(\ref{eq:LocalC}). In principle a coefficient $g>0$ of the cubic term can be included, i.e., $c^3\mapsto gc^3$, however here we choose $g=1$ for simplicity. The parameter $\Lambda$ relates to the conduction velocity (CV) restitution along the cable. In short, the CV, or wave front propagation speed with which a stimulus travels through tissue at a particular point depends on the local diastolic interval, i.e., the time spent depolarized after the previous stimulus has passed. CV depends on the diastolic interval through a nonlinear function $cv(d)$ that is typically monotonically increasing with but levels off for large $d$. In particular, since alternans cause the diastolic interval to vary along the cable, so does the CV, giving rise to an effect that is captured by the $\Lambda^{-1}\int_0^xe^{(x'-x)/\Lambda}a_n(x') dx'$ terms in Eqs.~(\ref{eq:MapC}) and (\ref{eq:MapA}), where $\Lambda=cv^2(d^*)/2cv'(d^*)$ and  $d^*$ is the diastolic interval at the onset of alternans. Typically $\Lambda$ is large so that $\Lambda^{-1}\ll1$. Here we have used $\Lambda=30$.

The parameter $\beta$ describes the effect of APD restitution, i.e., the dependence of voltage alternans on the voltage alternans at the previous beat. Because we are interested in calcium-driven alternans, $\beta$ should be chosen to be less than one. For simplicity, and since we have verified\cite{Skardal2014PRE} that this doesn't change the qualitative behavior of the model, we use $\beta=0$. The parameters $\alpha$ and $\gamma$ describe the bi-direction voltage$\to$calcium and calcium$\to$voltage coupling of the system. Here we have used $\alpha=\gamma=\sqrt{0.3}$.

Finally, the parameters $\xi$ and $w$ relate to the Green's function $G(x,x')=G(x'-x)+G(x'+x)+G(2L-x'-x)$ that appears in Eq.~(\ref{eq:MapA}) whose shape is described by Eq.~(\ref{eq:Kernel}). The parameter $\xi$ is a length scale that describes the width of $G(x)$, and physically indicates the length scale of electrotonic coupling, i.e., the spatial coupling that is due to the diffusion of voltage across tissue. The parameter $w$ is also a length scale, but has a different meaning. In particular, note that for $w=0$ the Green's function is simply a Gaussian kernel. Positive $w$ [which appears in only odd powers in Eq.~(\ref{eq:Kernel})] thus creates an asymmetry in the electrotonic coupling. Physically, this corresponds to the symmetry-breaking effect of a stimulus propagating in a given direction, here the positive direction from $x=0$ to $x=L$. Here we use $\xi=1$ and $w=0$ for simplicity.

\end{appendix}

\bibliographystyle{plain}

\end{document}